\documentclass[intlimits,twoside,a4paper]{article}

\usepackage{amsmath,amssymb}
\usepackage{graphicx}

\usepackage[T2A]{fontenc}
\usepackage[cp1251]{inputenc}

\usepackage[eqsecnum]{cmpj2}

\issue{2015}{18}{1}{13003}
\doinumber{10.5488/CMP.18.13003}

\title[What is liquid?  Symmetry: Lyapunov + Hamilton]{What is liquid?  Lyapunov instability reveals symmetry-breaking irreversibilities
hidden within Hamilton's many-body equations of motion}
\author[Wm.G. Hoover,  C.G. Hoover]{Wm.G. Hoover,  C.G. Hoover}
\address{
Ruby Valley Research Institute, Highway Contract 60, Box 601, Ruby Valley, Nevada 89833
}

\date{Received June 17, 2014, in final form November 25, 2014}

\begin{document}

\maketitle

\begin{abstract}
Typical Hamiltonian liquids display exponential ``Lyapunov instability'', also called
 ``sensitive dependence on initial conditions''.  Although  Hamilton's equations
are thoroughly time-reversible, the forward and backward Lyapunov  instabilities can differ,
qualitatively.  In numerical work, the expected forward/backward pairing of Lyapunov exponents
is also occasionally violated.  To illustrate, we consider many-body inelastic collisions in two
space dimensions. Two mirror-image colliding crystallites can either bounce, or not, giving
rise to a single liquid drop, or to several smaller droplets, depending upon the initial kinetic
energy and the interparticle forces.  The difference between the forward and backward
evolutionary instabilities of these problems can be correlated with dissipation and with the
Second Law of Thermodynamics.  Accordingly, these asymmetric stabilities of Hamilton's equations
can provide an ``Arrow of Time''.
  We illustrate these facts for two small crystallites colliding so as to make a warm liquid.
We use a specially-{\it symmetrized} form of Levesque and Verlet's bit-reversible Leapfrog integrator.
We analyze trajectories over millions of collisions with several equally-spaced time reversals.
\keywords Lyapunov instability, exponent pairing, chaotic dynamics, irreversibility
\pacs 05.10.-a, 05.45.-a
\end{abstract}

\section{Introduction}
\label{seq:1}

Doug Henderson and John Barker helped to set the stage for our own Nonequilibrium developments through
their equilibrium work on Thermodynamic Perturbation Theory \cite{b1}.  This novel approach solved the problem of
calculating accurate liquid-state thermodynamics by approximating the structure of a liquid with
hard-sphere or soft-sphere pair distribution functions. In our 2004 contribution, we described
Nonequilibrium Molecular Dynamics, the offshoot of classical mechanics designed to treat
mechanical and thermal gradients according to generalizations of Gibbs' statistical mechanics.
Since then, we have published a book summarizing these ideas \cite{b2}. Its successor, {\it Simulation and Control of Chaotic Nonequilibrium Systems}
has just been published by World Scientific Publishers in Singapore and released in March~2015.


\section{Lyapunov instability and Lyapunov spectra}
\label{seq:2}

A key finding of the nonequilibrium work was that steady-state distribution functions are singular
and fractal rather than Gibbsian and smooth, emphasizing the rarity of nonequilibrium states \cite{b3}.
In either of these cases, equilibrium or nonequilibrium, the necessary mixing in $n$-dimensional
phase space is facilitated by {\it Lyapunov instability}, the exponential growth of small
perturbations.  This instability is the focus of our present work.  Lyapunov instability is named
for a Russian, a gifted and prolific mathematician with roots in Saint Petersburg, Alexander
Lyapunov (1857--1918). Around 1979--1980, Shimada and Nagashima \cite{b4} as well as Benettin, Galgani,
Giorgilli, and Strelcyn \cite{b5} developed numerical methods for evaluating the spectrum of all $n$
Lyapunov exponents.

The spectrum describes the $n$-dimensional nature of Lyapunov instability in $n$-dimensional space.
The resulting orthogonal description of instabilities is much like the orthogonal description of
vibrations making up the solid-phase frequency distributions.
The basic idea is to follow the motion of $n$ satellite trajectories in the neighborhood of an
$n$-dimensional reference trajectory. The orthogonality of the $n$-dimensional vectors separating
the satellites from the reference can be enforced by Gram-Schmidt orthonormalization or by
an equivalent set of constraining Lagrange multipliers \cite{b6}.  When the motion is Lyapunov unstable, the
largest of the $n$ exponents describing the instability~--- $\lambda_1 \equiv \langle \ \lambda_1(t) \
\rangle $~--- the time-averaged rate at which {\it two} nearby trajectories separate~--- can be determined
from the growth rate of the first Lyapunov vector $\delta_1(t) \simeq \re^{\lambda_1 t}$.

Just as with the whole spectrum, this determination of $\lambda_1(t)$ can be done
in either of two ways: (i) rescale the distance between a satellite trajectory and the reference
trajectory at each discrete timestep or, (ii) constrain the length of the offset vector $|\delta_1|$
with a Lagrange multiplier, $\lambda_1(t)$.  The Lagrange multiplier approach \cite{b6} entails
one multiplier for each of the $n(n-1)/2$ angles defined by a pair of vectors, plus $n$ additional
multipliers for the lengths of the vectors.  Since the constrained problem, along with all of its
Lagrange multipliers, is {\it time-reversible}, the complete set of multipliers going forward
needs only to change sign to maintain the orthonormality constraints in the reversed direction.
Numerical work shows that this reversibility is illusory (as is quite well known to the experts).
The reversed set of vectors is unstable, as we will see presently.

\begin{figure}[!b]
\centering
\includegraphics[height=0.5\textwidth,angle=90.]{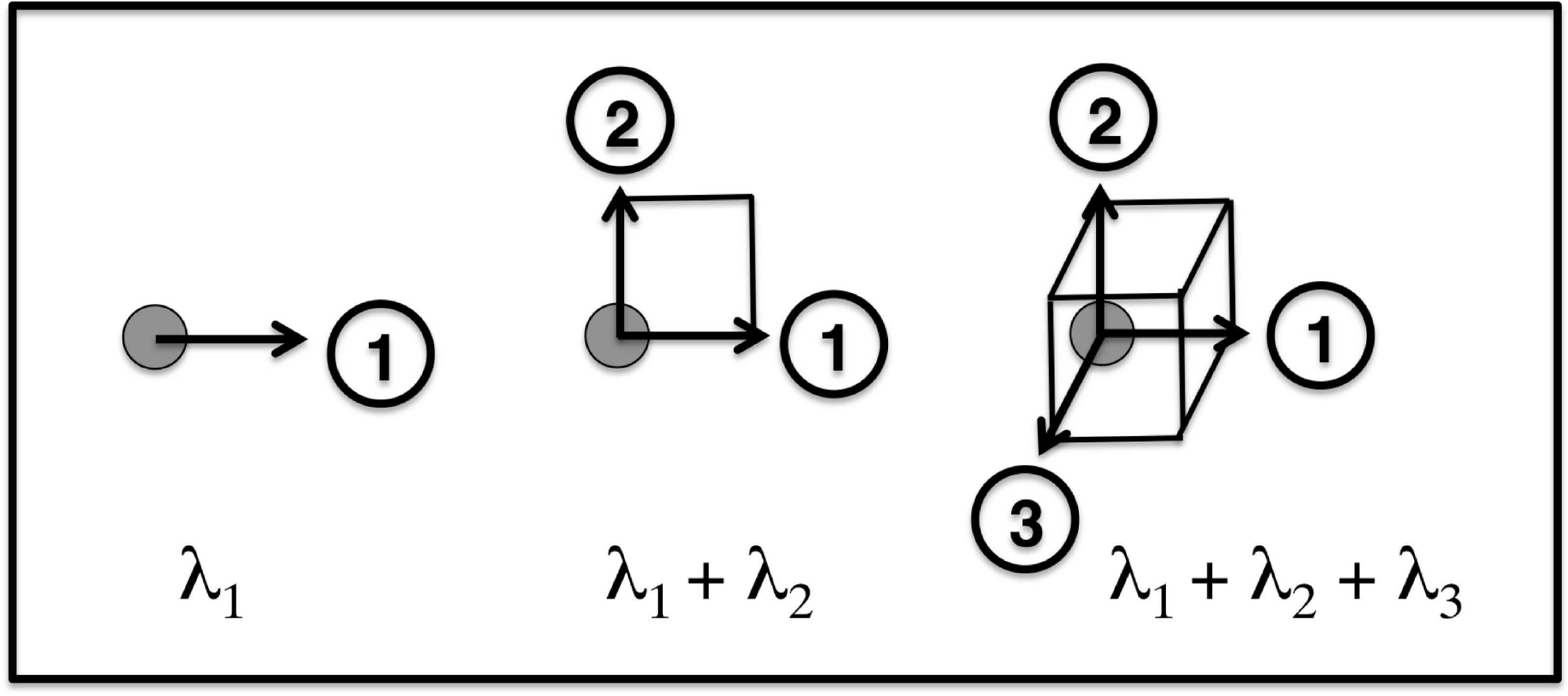}
\caption{
The Lyapunov exponents $\{\lambda_1,\lambda_2,\lambda_3\}$ are respectively the growth rates
$(\dot \delta/\delta)$ of small orthogonal vectors $\{\delta_1,\delta_2,\delta_3\}$ in one-,
two-, three-dimensional subspaces of $n$-dimensional phase space.  If the vectors are allowed to
grow during each timestep, then they are {\it rescaled} in length by the Gram-Schmidt procedure, which
also maintains their orthogonality.  If the vectors are instead constrained to a constant length
with Lagrange multipliers$\{\lambda_1,\lambda_2,\lambda_3\}$, those multipliers are identical
to the ``local'' (time-dependent) Lyapunov exponents \cite{b6}.
\label{fig1}}
\end{figure}

A second Lyapunov exponent, $\lambda_2(t)$, can be added to the first to describe the rate
at which a two-dimensional {\it area} in $n$ space changes with time.  Continuing this process
to the third, fourth,~\ldots multiplier, the sum of {\it all} $n$ Lyapunov exponents gives the rate
at which the $n$-dimensional hypervolume in the $(q,p)$ phase space changes with time:
\[
\dot \otimes(t)/\otimes(t) \equiv \sum_1^n \lambda_i(t) \ .
\]
Figure~\ref{fig1} illustrates the relationship of the Lyapunov exponents to the orthogonalized vectors
separating the $n$ satellite trajectories from the $n$-dimensional reference trajectory.

For simplicity, we develop all of our manybody models in two space dimensions, using particles
of unit mass.  The dimensionality of the corresponding phase space is four times the number of
particles, $n = 4N$.  There is a separate phase-space direction for each particle coordinate
and momentum (velocity, for particles of unit mass):
\[
\{ q,p  \}  \equiv  \{  x_i,  \dot x_i, y_i,  \dot y_i  \}  .
\]
Typically, in many-body systems, the Lyapunov exponents are of the same order as the collision rate,
and the ``spectra'' of all the exponents resemble the Debye spectra of solid-state physics.

\begin{figure}[!b]
\centering
\includegraphics[height=0.4\textwidth,angle=90.]{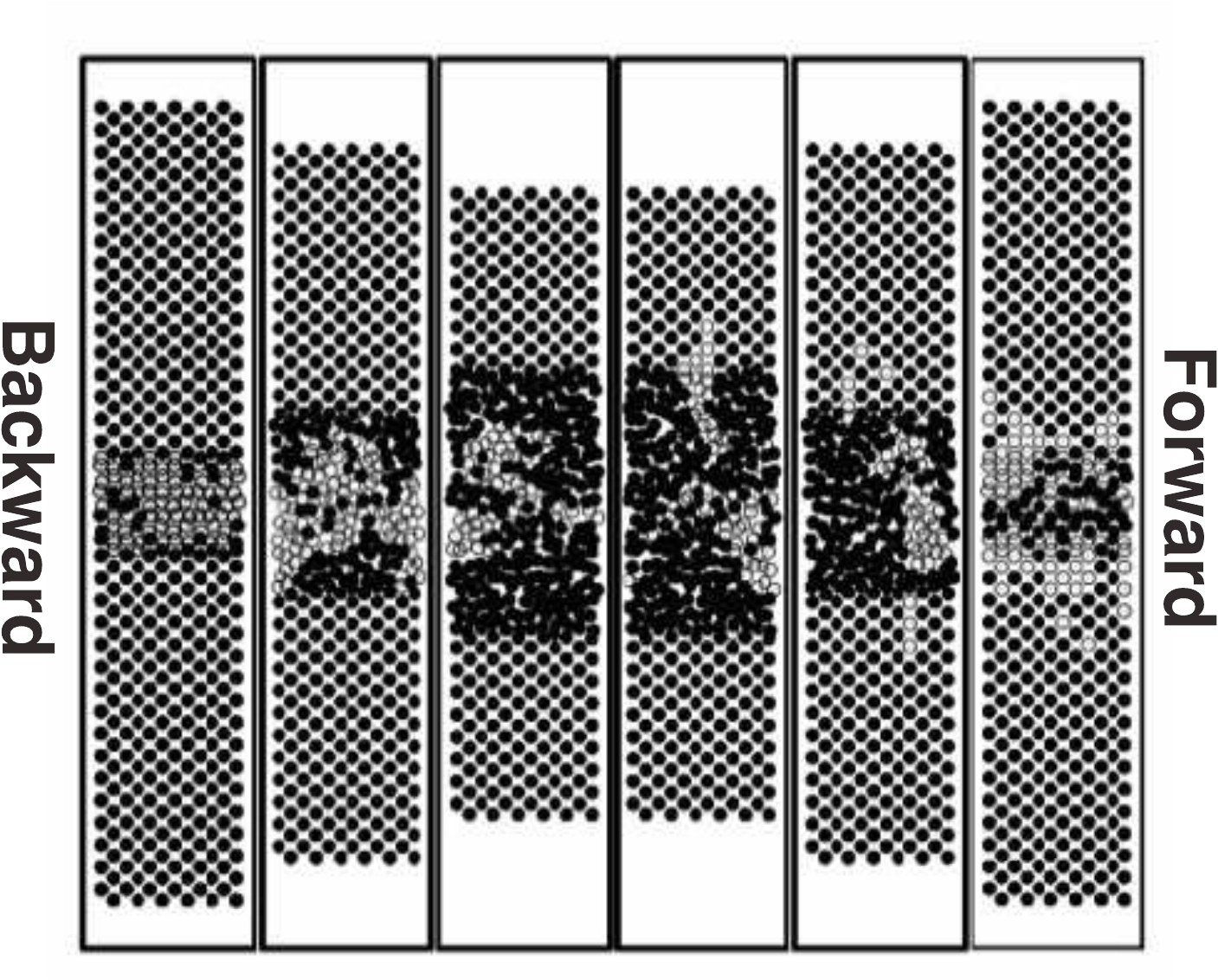}
\caption{
Important particles, shown as open circles, in the collision of two periodic 240-particle
triangular-lattice crystallites.  The forward and backward times are 2, 4, and 6, with
reversal at 12.  See reference \cite{b7} for more details, especially figures~20
and 21 of that reference.
\label{fig2}}
\end{figure}

The time-reversibility of Hamilton's equations of motion extends also to the time-reversibility
of the differential equations governing the Lyapunov vectors $\{  \delta(t)  \}$ and their associated
exponents $\{  \lambda(t)  \}$.  This has two interesting consequences:
(i) Directly from Hamilton's equations of motion one would (na\"ively) expect that for every
positive exponent and vector there is a time-reversed pair, with all the momenta reversed:
\[
\big\{ +\lambda_i(+\delta q,+\delta p )_t \longleftrightarrow  -\lambda_{n+1-i}( +\delta q,-\delta p )_t  \big\}  .
\]
Although this is true, it turns out that only one of the vectors in each pair is ``stable''.  The
{\it observed} vectors going forward in time are quite distinct from those going backward.
(ii) For every Lyapunov vector of the form $\delta = (\delta q,\delta p)$ there is also another
{\it paired} orthogonal vector with an oppositely-signed Lyapunov exponent and with the coordinate and
velocity components of the original vector switched:
\[
\big\{  +\lambda_i(+\delta q,+\delta p )_t \longleftrightarrow  -\lambda_{n+1-i}( -\delta p,+\delta q )_t  \big\} .
\]
By permuting the components of the vectors in this way, orthogonality is guaranteed.  This pairing
is {\it mostly} true.  Typically, there {\it is} a simple relationship between vectors corresponding to Lyapunov
exponents with opposite signs.  But we will see that this is not always the case.  The occasional
exceptions brought about the present work.

We previously investigated the first and simpler of the two pairing ideas mentioned above, comparing the
forward and backward phase-space offset vectors $\delta_1^\textrm{f}$ and $\delta_1^\textrm{b}$ for planar shockwaves
generated by two colliding crystals.  These mirror-image crystals moved toward each other in the $x$
direction.  In the $y$ direction, the boundary conditions were periodic \cite{b7}.  See  figure~\ref{fig2}.
{\it Forward} in time, the ``important particles'', making above-average contributions to $\delta_1^\textrm{f}$,
were concentrated within the hot shocked material.  {\it Backward} in time, and with the very same
configurations with opposite velocities, the above-average contributions were less spatially
concentrated \cite{b7}.

In order to eliminate transients in such calculations, we hit upon the idea of cycling a {\it bit-reversible}
(exactly reversible, to the last bit) simulation forward and backward in time until the forward and
backward vectors $\delta_1^\textrm{f}$ and $\delta_1^\textrm{b}$ had converged to machine accuracy.  Again, the important
particles going forward and backward in time were qualitatively different \cite{b8,b8'}.   Similar effects
were found for binary collisions of two crystallites in the absence of any spatial periodicity \cite{b9}.
All these simulations, with or without spatial or temporal periodicities,  agreed in finding qualitative
differences between the Lyapunov vectors forward and backward in time. In reference \cite{b9}, where the full
Lyapunov spectrum for two colliding 37-particle hexagons was computed, the only vector and exponent pairing
observed was quite imperfect.  Though these calculations were bit-reversible, they spanned only tens of
thousands of timesteps.  We consider much longer simulations in the present work.

\begin{figure}[!t]
\centering
\includegraphics[height=0.7\textwidth,angle=90.]{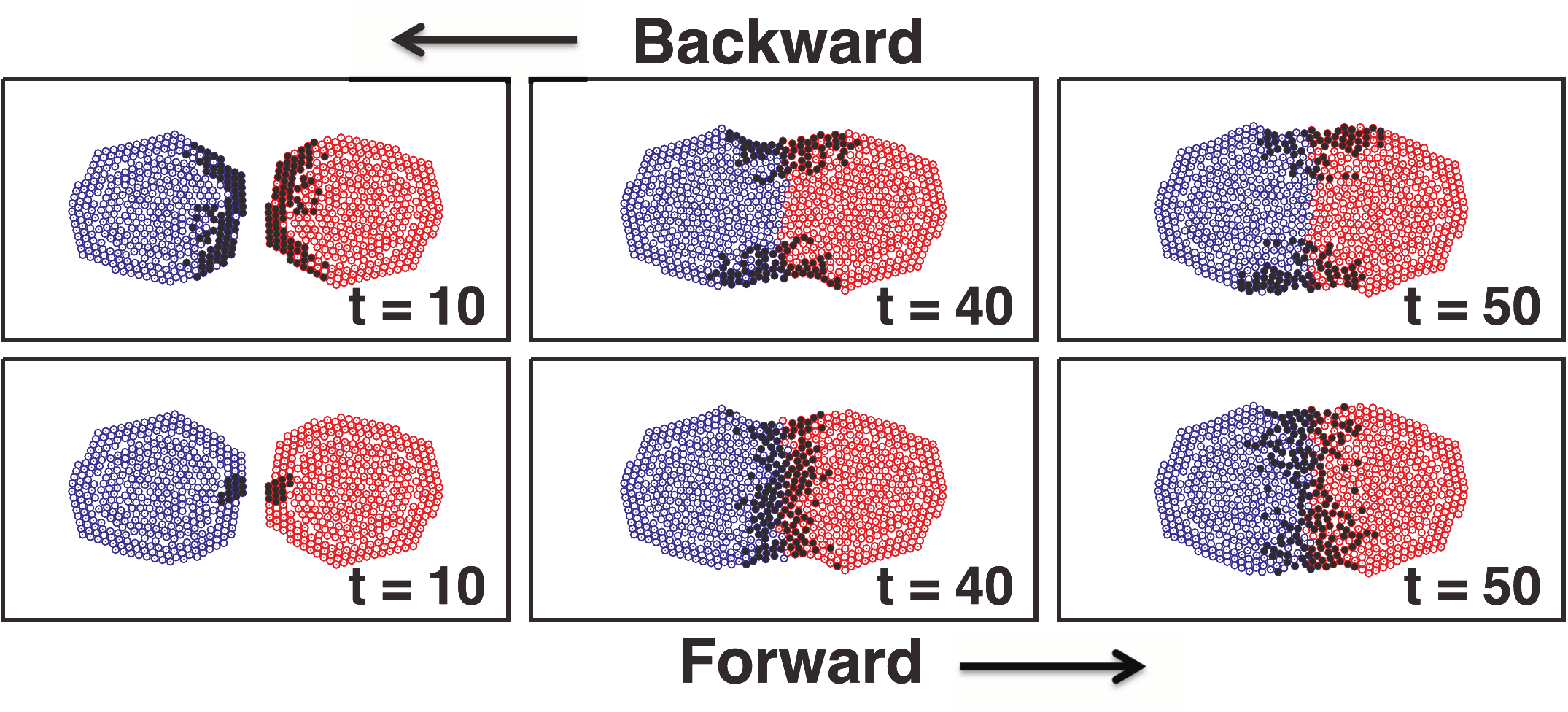}
\caption{(Color online)
Important particles, shown in black, in the collision of two 400-particle balls reversed at time
$t=100$.  The snapshots correspond to three different times with identical particle coordinates
in both the forward and the backward directions of time.  The $\delta$ vectors giving the forward
and backward values of $\lambda_1$ are quite different in the two time directions.  See figures~12 and 13 of reference~\cite{b9}.
}\label{fig3}
\end{figure}

The leading vectors going forward in time emphasize the leading edges of the crystals, where the
collision is taking place.  In the absence of periodic boundary conditions, the leading vectors going
backward in time (and we will soon describe the best way to go backward) instead emphasize the
``necking'' region, where the compound liquid drop formed by the colliding crystals relives its past
history as two separate bodies.  See figure~\ref{fig3} for the collision of two 400-particle crystalline
balls.  In that figure, the particles making above-average contributions to the largest Lyapunov exponent
going both forward and backward in time are emphasized.

In the present work, we reduce the intricacies of the Lyapunov spectrum and the instabilities it
describes by considering smaller systems for longer times.  These are all Hamiltonian systems with
two crystallites colliding to form one or more fragments.  These smaller simpler systems make it possible to
study Lyapunov instability and the pairing of vectors with greater precision.

\begin{figure}[!b]
\centering
\includegraphics[height=0.45\textwidth,angle=90.]{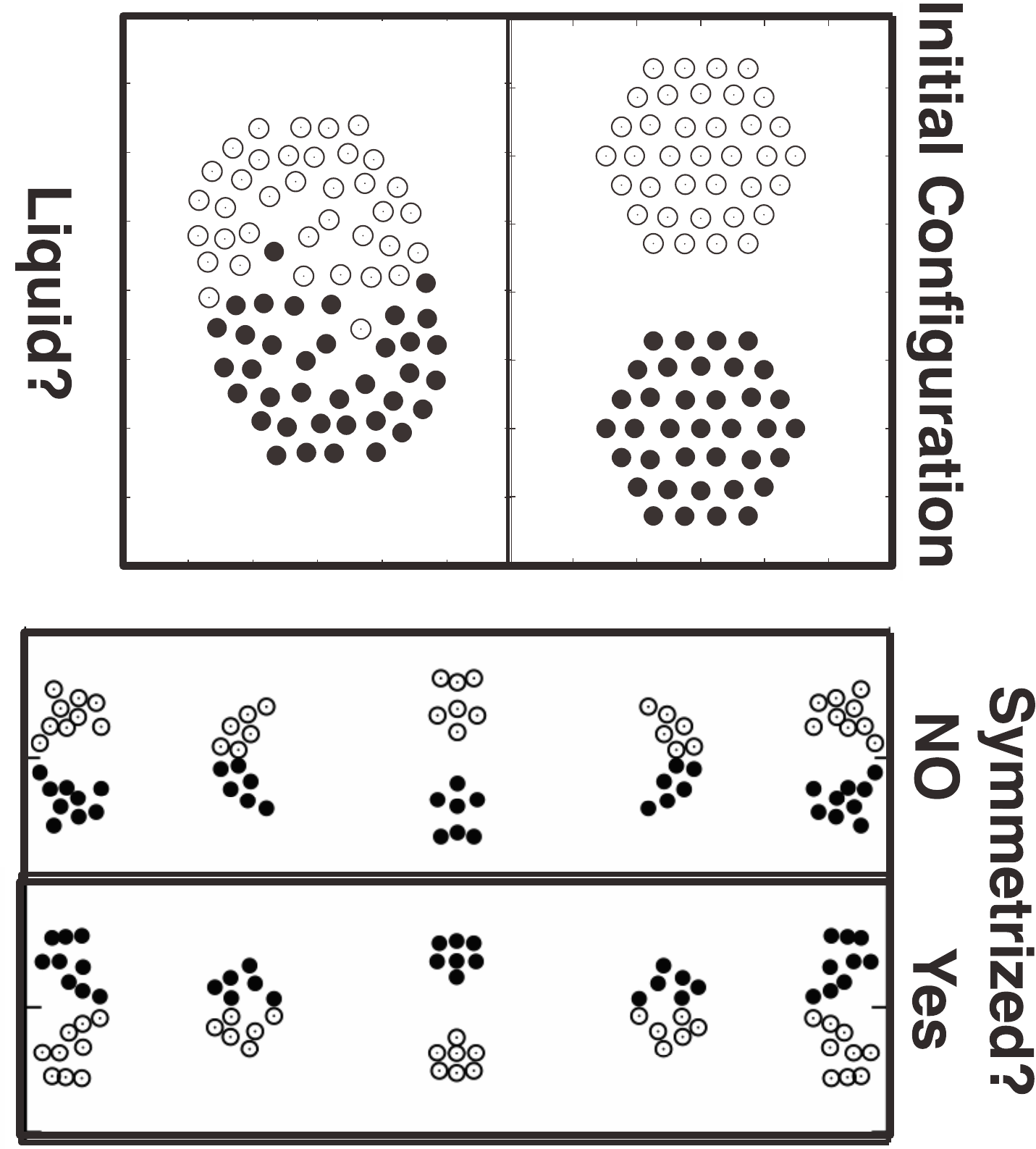} 
\caption{
Pairs of 37-particle crystallites and their collision products. The initial condition for all of these
simulations corresponds to the upper left-hand illustration.  At the low velocity of $\pm 0.10$, the
result is the ``liquid'' ball shown at the lower left.  At the higher velocity of $\pm 0.50$,
 the results for bit-reversible simulations with and without symmetrization (described in section~\ref{seq:6})
 are shown at the right where the 74 particles have separated into six fragments.  The
two bit-reversible simulations differ as a result of the Lyapunov-unstable amplification of computer
roundoff errors, as discussed in section~\ref{seq:6}.
}\label{fig4}
\end{figure}

Figure~\ref{fig4} shows sample 74-particle snapshots for two different initial velocities. At relatively
low velocities, the colliding hexagons can bounce or coalesce.  At higher velocities, several
smaller crystallites or drops are formed.  To simplify both the dynamics and the analysis for corresponding
pairs of particles in the two 37-particle hexagons, we choose inversion-symmetric initial conditions:
\[
\big\{  x_{\rm Left} + x_{\rm Right} = y_{\rm Left} + y_{\rm Right} = 0 ; \quad
\dot x_{\rm Left} + \dot x_{\rm Right} = \dot y_{\rm Left} + \dot y_{\rm Right} = 0 \big\}  .
\]
In order to propagate the particles reversibly, we use Levesque and Verlet's bit-reversible
algorithm \cite{b10} which we detail in the following section.

\section{Levesque-Verlet bit-reversible simulations}
\label{seq:3}

The study of Lyapunov instabilities requires special numerical methods.  Since our aim here is to
compare the stabilities of forward and backward motion equations for {\it millions} of timesteps, we
begin  with Levesque and Verlet's observation that the ``Leapfrog'' algorithm for solving Newton's
equations of motion can be made precisely time-reversible by restricting the particle coordinates to
(large) integer values:
\[
\left\{  q_{t + dt} \equiv 2q_t - q_{t - dt} + \left[  (\rd t^2/m)F(q_t) \right]_{\rm int}  \right\} \longleftrightarrow
\left\{  q_{t - dt} \equiv 2q_t - q_{t + dt} + \left[ (\rd t^2/m)F(q_t) \right]_{\rm int}  \right\}.
\]
Apparently, this identity ensures reversibility. All that has to be done is to compute and round off
the force terms, as indicated by the brackets $[  \ldots  ]_{\rm int}$, in precisely the same way
whether going forward in time, to $t+\rd t$, or backward, to $t-\rd t$.  By using ``long'' 16-digit
integers, a satisfactory precision can be obtained.  An alternative is to store the billions of
trajectory coordinates describing the forward trajectory.  With either method, a strictly
``bit-reversible'' reference trajectory can be generated forward in time and can then be used in
reversed order to describe its backward relative.  Since the bit-reversible Leapfrog technique relies
on a ``conservative'' one-to-one mapping of successive pairs of configurations, it cannot be applied to
``dissipative'' motion equations.  Dissipative motions cause the phase volume to shrink.  Ultimately,
that chaotic shrinking generates fractal strange attractors \cite{b2,b3,b7}.

The application of the bit-reversible algorithm to the computation of Lyapunov spectra was pioneered by
M.~Romero-Bastida, D.~Paz\'o, J.M.~L\'opez, and M.A.~Rodr\'iguez \cite{b11}. With a strictly reversible
reference trajectory, the motion of the $n$ nearby satellite trajectories can then be generated with
straightforward Runge-Kutta integration.  The corresponding numerical method is described in section~\ref{seq:4}.
These ideas are then applied to the idealized liquid Hamiltonian described in section~\ref{seq:5}.  The
results, for typical collisions, are detailed in section~\ref{seq:6}.  The final section~\ref{seq:7} sums up the
connection of these differences to irreversibility, as described by the Second Law of Thermodynamics.

\section{Combining Newtonian and Hamiltonian mechanics }
\label{seq:4}

In our molecular dynamics work we combine the Newtonian and the Hamiltonian forms of mechanics.
The coordinate-based Newtonian Leapfrog algorithm advances {\it pairs} of coordinate configurations
$\{  q_t,q_{t\pm \rd t}  \}$ while Hamilton's first-order motion equations advance from one
$\{ q_t,p_{t}  \}$ phase point to the next $\{  q_{t+\rd t},p_{t+\rd t}  \}$:
\[
\big\{ \ddot q = F(q)  \big\} \qquad \text{\it versus} \qquad \big\{  \dot q = p ; \quad \dot p = F(q)  \big\}  .
\]
Combining the two forms of mechanics requires a {\it definition} of momentum based on the coordinate
information generated by the Leapfrog algorithm. The unimaginative first-order choice, with errors
\linebreak $(1/2)\stackrel{...}{q}\rd t^2$,
\[
p_t \simeq \left[ q_{t+\rd t} - q_{t-\rd t} \right]/(2\rd t) ,
\]
can and should be improved upon by using instead the third-order definition \cite{b9},
\[
p_t \equiv (4/3)\left[ q_{t+\rd t} - q_{t-\rd t} \right]/(2\rd t) - (1/3)\left[ q_{t+2\rd t} - q_{t-2\rd t} \right]/(4\rd t) .
\]
The formal error in this last definition is $-(1/30)\stackrel{.....}{q}\rd t^4$.

\section{Numerical liquid models for the collision process}
\label{seq:5}

\begin{figure}[!t]
\centering
\includegraphics[height=0.6\textwidth,angle=90.]{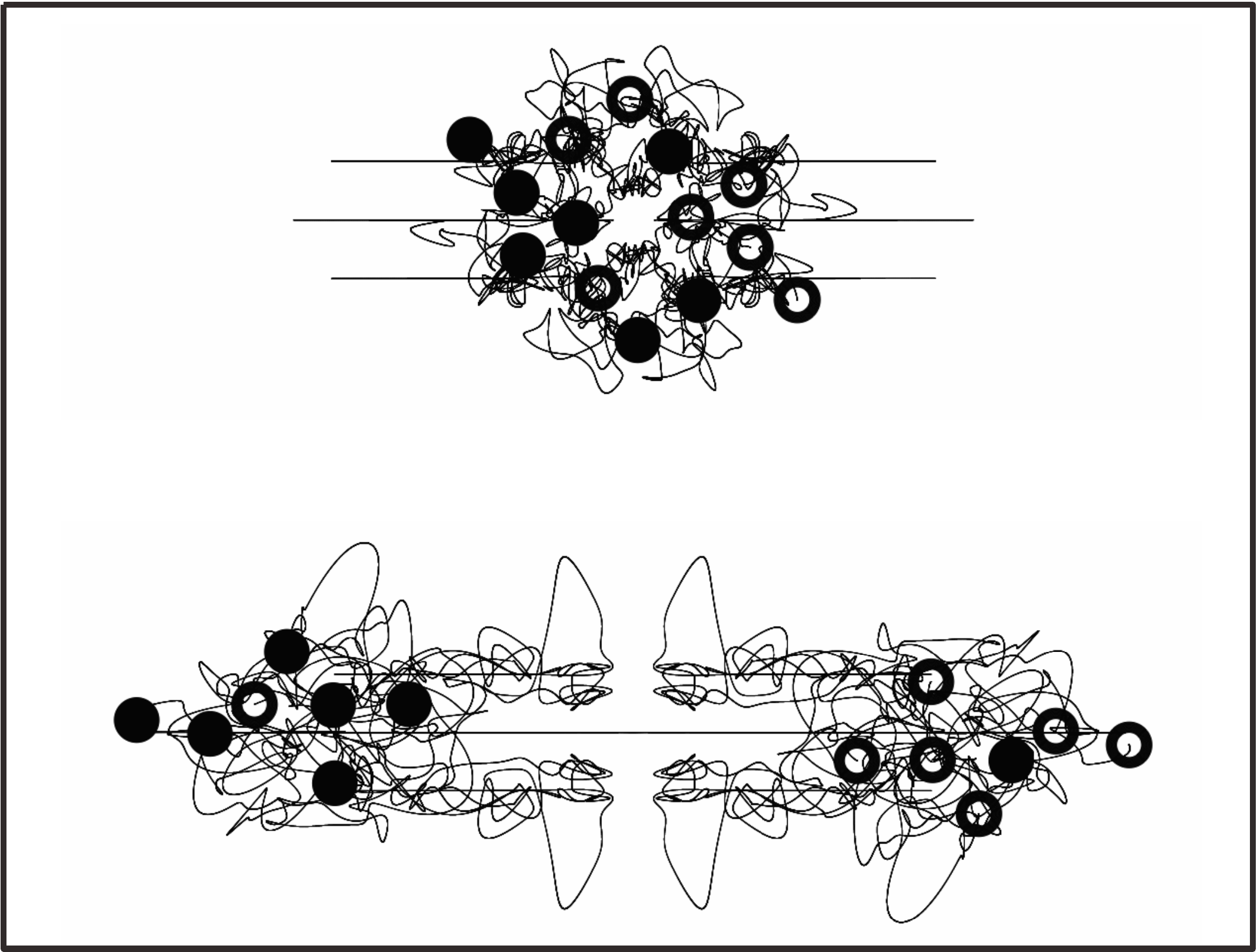} 
\caption{
Pairs of 7-particle crystallites and their collision products for two choices of the initial
velocities, $\pm 0.1$ (above) and $\pm 0.5$ (below).  Notice that in the higher-speed
collision, the crystallites exchange one particle before separating. The inversion symmetry has
been maintained by {\it symmetrization} at each timestep but the original fourfold symmetry of
the hexagons has been lost.  The symmetrization process is described in section~\ref{seq:6}.
}\label{fig5}
\end{figure}

To minimize numerical errors, it is useful to choose very smooth force laws.  As demonstration
problems we consider here collisions of two hexagonal crystallites. Snapshots appear in
 figure~\ref{fig4} and \ref{fig5} for collisions of both 37-particle and 7-particle crystallites.  We follow
reference \cite{b9} and use a many-body attractive binding energy, $(1/2)(\rho_i - 1)^2$ for each
Particle $i$.  This potential, when differentiated gives the force on Particle $i$ due to Particle
$j$ as $(2 - \rho_i - \rho_j)\nabla_iw_{ij}$.  Each Particle's personal density is computed using
Lucy's weight function \cite{b2}. {\it All} particles $\{ j \}$  (including $i$) within a distance $h$ make a contribution to the density there:
\[
\rho_i \equiv \sum_{j=1}^{14 \ {\rm or} \ 74}w(r<h) , \qquad w = (5/h^2\pi)(1+3z)(1 - z)^3  , \qquad
z \equiv r/h .
\]
The normalization constant $(5/h^2\pi)$ is chosen so that the integral of the weight function
is unity:
\[
\int_0^hw(r)2\pi r\rd r \equiv 1 .
\]
For the model systems discussed here, with $N=2\times 37=74$ or $N=2\times 7=14$, we have used
$h=3.00$ and $h=3.50$, respectively.  A typical nearest-neighbor distance in all of these problems
is of order unity.

In order to minimize the occurrence of very closely-spaced pairs of particles, it is
expedient to include a short-ranged repulsive potential in the Hamiltonian.  For our
demonstration problems we choose a very smooth ``soft-disk'' short-ranged pair potential:
\[
\phi_{\rm pair}(r<1) = (1 - r^2)^4  .
\]
With these specially smooth choices for the attractive binding potential and the repulsive pair potential
(and with our third-order definition of the momentum), the energy throughout a run remains constant with
six-figure accuracy.

The satellite solutions for each timestep are launched basing the offset vectors on the
current leapfrog values of the coordinates and their associated momenta  $\{ q,p  \}$.
Then, a fourth- or fifth-order Runge-Kutta integrator provides close to machine accuracy in
incrementing the satellite motions with a timestep of~0.001.

\section{Crystallite collisions and their time-reversed twins}
\label{seq:6}

\begin{figure}[!h]
\centering
\includegraphics[height=0.5\textwidth,angle=90.]{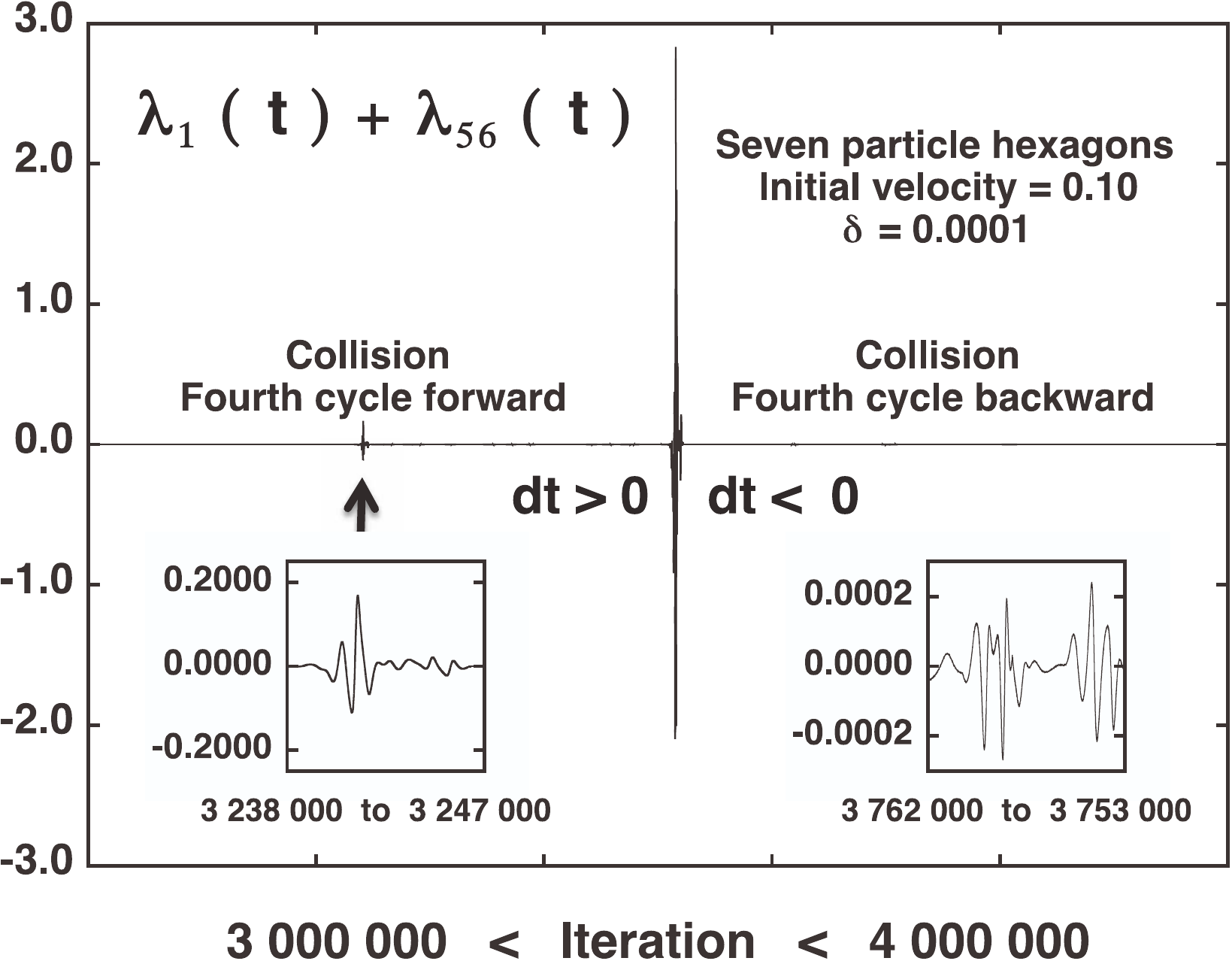} 
\caption{
Deviation from exponent pairing going forward in time (left of center) and backward (right
of center).  The Lyapunov sum, $\lambda_1 + \lambda_{56}$ is plotted in both time directions
for a $7+7$ particle collision with velocities $\pm 0.10$.  The data are plotted for the third
repetition, for iterations between 3 million and 4 million, where $\rd t = \pm 0.001$. The deviation
from pairing going forward in time is at its maximum at a time near 243, corresponding to iteration
numbers from $3\, 238\, 000$ to $3\, 247 \, 000$.  The insets show details for time windows of width 9.  In
the reversed motion with $\rd t$ negative and at time 243, the pairing is nearly perfect, invisible
on the main plot.  The reversed-time deviation from pairing is {\it three orders of magnitude smaller}
than the forward-time deviation.  The huge backward spike near iteration $3\, 510\, 000$ corresponds to the
conversion of the forward set of $\delta$ vectors into the backward set following time
reflection at iteration $3\, 500\, 000$ where $t=500$.
}\label{fig6}
\end{figure}

Continuing to pursue simplicity we choose initial conditions with {\it symmetric} coordinates
and momenta for the two colliding crystallites.  Sample geometries are shown in figures~\ref{fig4} and \ref{fig5}.
\[
\big\{ q_i = - q_{N+1-i} ; \quad p_i = -p_{N+1-i} \big\}, \qquad i = 1 \dots N .
\]
Figure~\ref{fig6} shows the evolution of the summed-up largest and smallest Lyapunov exponents, $\lambda_1(t) + \lambda_{56}(t)$, both forward and backward in time, but for 14 particles rather than 74 or 800 and for
one million timesteps, half a million forward and half a million reversed. The initial orientations of
the offset vectors, $\left\{\delta \equiv (q,p)_{\rm sat} - (q,p)_{\rm ref} \right\}$, can either be chosen ``randomly'' or as rows of a unit matrix.  A convenient length for the vectors in these numerical
simulations is 0.0001. For our collision problems it takes about 50 000 timesteps (with $\rd t = 0.001$)
for the time-dependent vectors to converge stably to a set independent of the initial conditions.

We reverse the direction of time periodically, by changing the sign of $\rd t$ every half million
timesteps, with nineteen changes in all in the course of a ten-million-timestep run.  The reference
trajectories
for the forward and backward segments are each repeated ten times and are identical to machine
accuracy using the Levesque-Verlet bit-reversible integrator. The satellite trajectories rotate
about the reference trajectory, constrained to remain orthogonal and of constant length. After
the first repeated forward and backward segments, these trajectories hardly change in subsequent
segments.  The only significant changes occur near the middle of the collision process, forward in
time.  For typical details see the inset of figure~\ref{fig6}.

One would expect that the left-right and up-down symmetries imposed on the initial conditions
would persist throughout any bit-reversible calculation.  Wrong!  Because the {\it order} in
which the summed up forces give the total force on a particle are not necessarily symmetric,
the sums can, and eventually do, differ in the last decimal, with the resulting difference
amplifying exponentially in time.

Take a toy-model example in which Particles 1, 2, 3, and 4 are arranged in order on a line
with the pair interactions added up in the usual $(i < j)$ order.  The force on the leftmost
particle, Particle 1, is a sum of first, second, and third neighbor forces, in that order.
The force on the rightmost particle, Particle~4, is instead the sum of third, second, and
first neighbor forces, the opposite order.  On a finite-precision machine, the totals are
likely different.  Since the average force is typically close to zero, it is often the
case that significant figures are lost in the process of summing the forces on particles.

\begin{figure}[!b]
\centering
\includegraphics[height=0.45\textwidth,angle=90.]{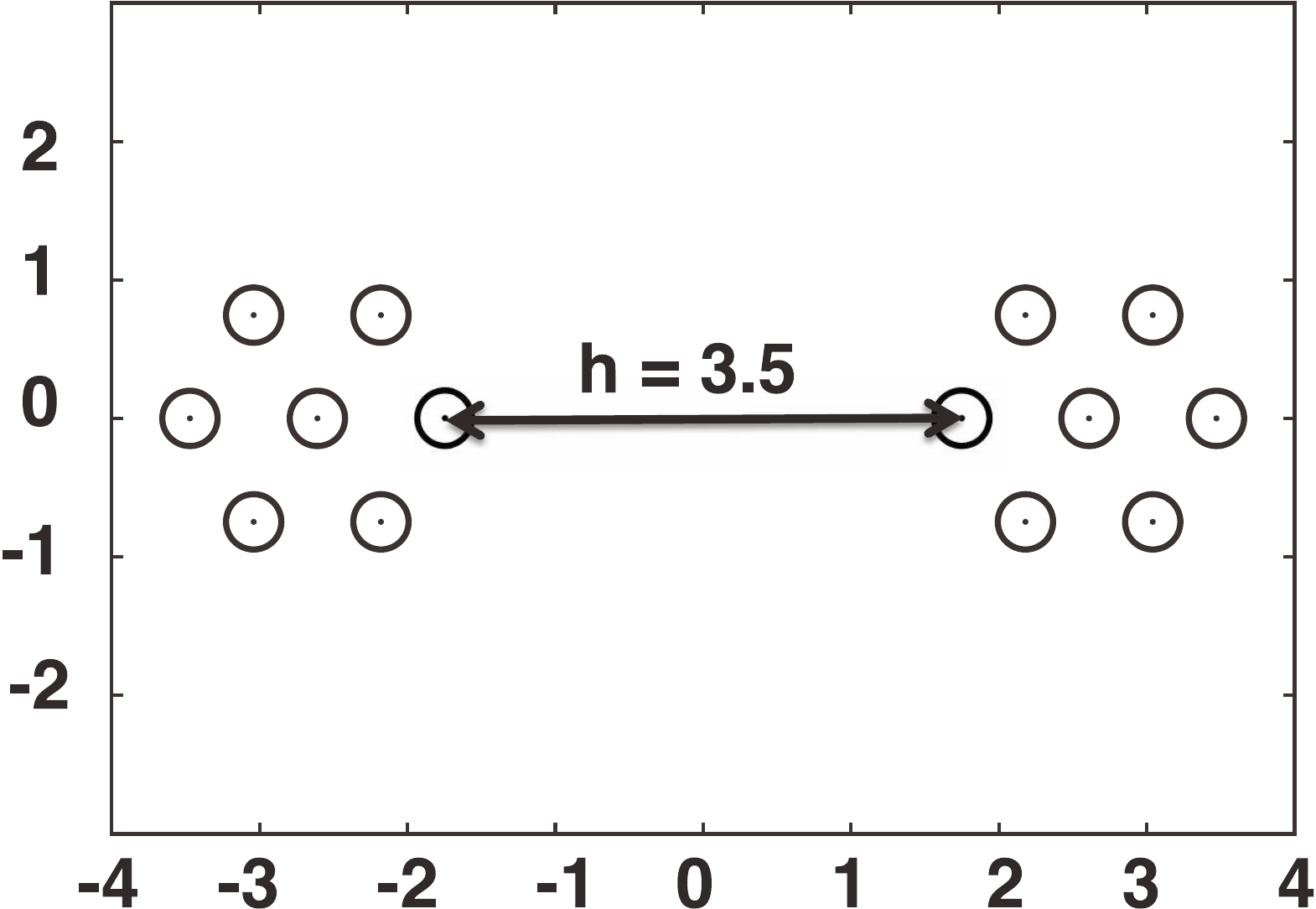} 
\caption{
Two 7-particle cold hexagons at the moment of collision, where the separation of their leading
edges is $h = 3.5$.  The original center-to-center distance at
time zero was 50.  This system was followed for $10\,000\,000$ timesteps alternating segments with
$500\,000$ forward steps followed by $500\,000$ backward steps in order to ensure the convergence of
the offset vectors both forward and backward in time.
}\label{fig7}
\end{figure}

It is annoying to learn that an initially perfectly symmetrized bit-reversible problem loses its symmetry
in a macroscopic way after approximately $50\, 000$ timesteps, about the same time as is required
for Lyapunov exponent pairs to pair up. Just as in the simple four-particle toy-model example,
the symmetrized problem loses its symmetry due to the capricious nature of the ordering of
the force sums.  Lyapunov instability can quickly move these tiny errors from the last decimal
place to the first, magnifying them by a factor of $10^{16}$, and so producing configurations
which visibly lack symmetry.  For an inadvertent example see the asymmetry shown in the
bit-reversible figure~5 of reference \cite{b9}.  In our ``symmetric'' demonstration problems here, the
inversion symmetry {\it can} be maintained by the expedient of {\it symmetrizing} the forces on each
member of the particle pairs at every timestep:
\[
F_i = [ F_i - F_{i+7}  ]/2; \qquad F_{i+7} = [ F_{i+7} - F_i ]/2 ,
\]
likewise taking care to compute symmetrized densities for each particle:
\[
\rho_i = [ \rho_i + \rho_{i+7} ]/2; \qquad \rho_{i+7} = [  \rho_{i+7} + \rho_i  ]/2 .
\]

The {\it pairing} of local (instantaneous) Lyapunov exponents has been much
discussed \cite{b12,b13,b14,b15,b16,b17}, mostly for very small systems with only a few degrees
of freedom. Our 74-particle simulations indicated exponent pairing {\it most} of the time.  But
the complexity of the Gram-Schmidt calculations, orthogonalizing 296 vectors at each timestep,
left unclear the reason for the occasional loss of pairing.  A numerical source seemed
likely because the largest Lyapunov exponent was closely reproduced (visually there was
no detectable change) from one million-timestep sequence to the next while variations in
the smallest exponent, $\lambda_{296}$ were visible and wholly responsible for the nonzero
values of the sum, $\lambda_1(t) + \lambda_{296}(t)$. To avoid any uncertainty, we chose to
concentrate on the simpler 14-particle problems with their reduced demands on the Gram-Schmidt
orthonormalization step.

The moment of collision for the problem that we choose to study in detail is shown in  figure~\ref{fig7}. Originally, at time zero, the center-to-center separation of the two crystallites was 50 with
all particle velocities $(0,+0.1)$ in the leftmost hexagon and $(0,-0.1)$ in the rightmost
hexagon.  The least-energy nearest-neighbor separation is $0.8611\, 2127\, 0463$, which minimizes the
potential energy, and there is no thermal motion.  Without any interaction, the two crystallites
would have overlapped perfectly at a time of 250.  As shown in the figure, the collision actually
begins at time $[ 50-3.5-2 \times 0.8611 ]/0.2 \simeq 223.9$.  By a time of 250, the
collision has converted two cold crystallites to a single warm drop. All of the results
described here for this collision use an offset vector length $|  \delta  | = 0.0001$ and a timestep $\rd t = 0.001$.

Figure~\ref{fig6} showed the gross features of the sum of the first and last Lyapunov exponents over the
fourth interval of one million steps.  On the scale shown, all of the intervals from the second
through the tenth behaved identically, with the Lyapunov exponents paired during the reversed motion
but consistently undergoing an episode of nonpairing, going forward in time, at a time of 235
(corresponding to $235\, 000$, $1235\, 000$, $2\,235\, 000$, {\ldots} timesteps).  A
detailed investigation shows that the lack of pairing occurs during a shearing motion of the central
rows of particles relative to the upper and lower ones.  It is this same shearing motion which was
associated with the dynamics of hard disks at melting in 1963 \cite{b18}.  For most of the simulation,
the first and the last exponents nearly sum to zero, $\lambda_1 + \lambda_{56} \simeq 0$.  The second
half of the run retraces the configurations beginning at the maximum time of 500.
Figure~\ref{fig6}
shows that $\lambda _1$ reverses visually to a time of 488 where the set of $\delta $ vectors begins
its change from the unstable reversed forward vectors to the stable backward ones.  For the remainder
of the run, back to the initial configuration at time zero, no further disturbances to pairing are
observed.  Indeed, as the figure shows, the pairing is mostly quite good (as the sum is very close to zero)
forward in time too, in the  interval $ 0 < t < 500 $, except where the collisional effects are a
maximum near $t=243$.  It is interesting to see that both before and after the collision, the pairing
is nearly perfect.  Within the collision, there are some brief but quite significant differences.

\begin{figure}[!t]
\centering
\includegraphics[height=0.5\textwidth,angle=-90.]{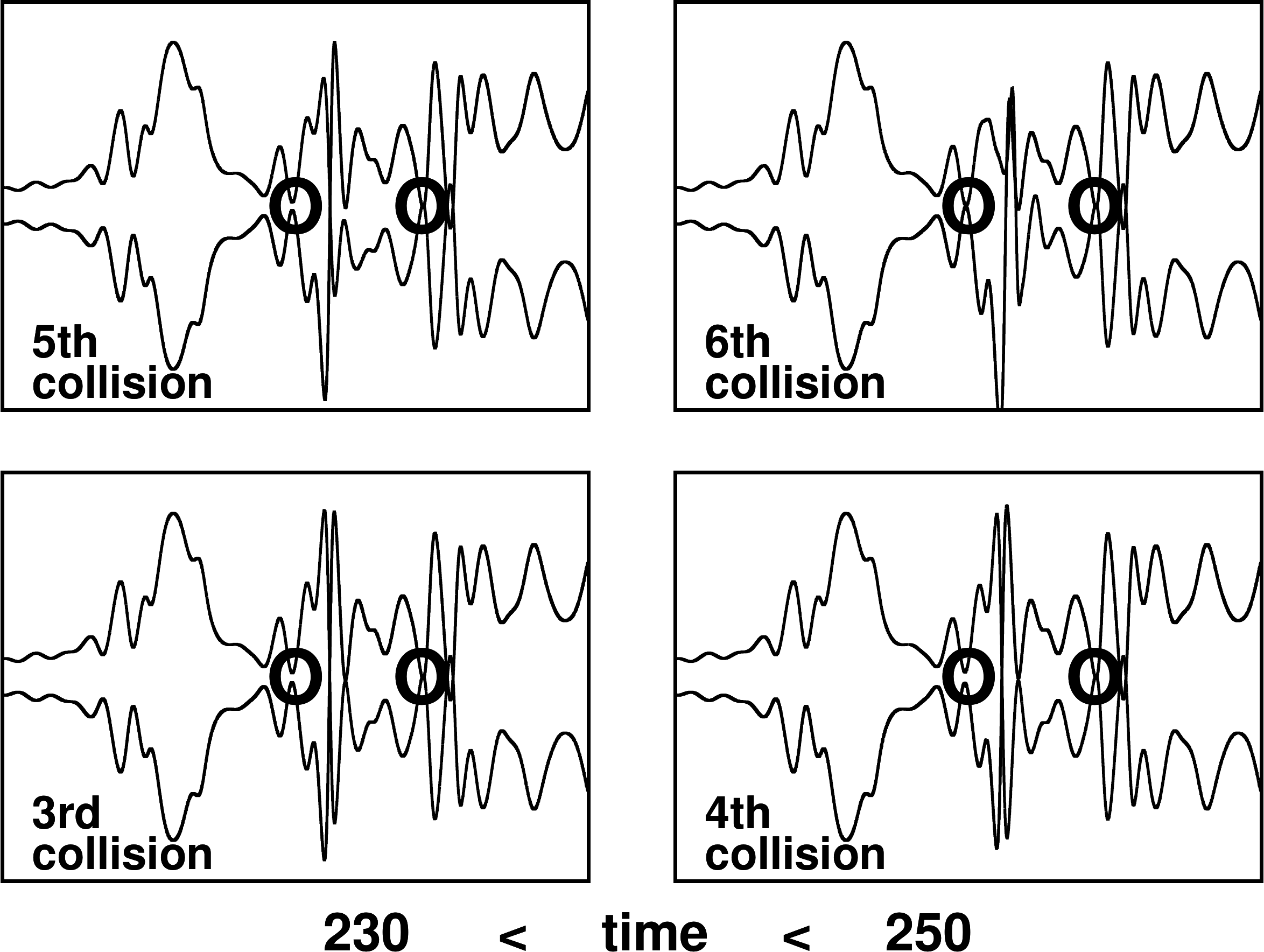} 
\caption{
Transition region details are illustrated for four different collisions, separated by one million
timesteps.  We show the time-dependence of the first and last Lyapunov exponents,
$+\lambda_1(t) \simeq -\lambda_{56}(t)$ for the inelastic collision of two 7-particle hexagons with
opposite initial velocities $(\pm 0.1,0.0)$.  Apart from the transition region between the two
open circles, the pairing of the exponents is nearly perfect.
}\label{fig8}
\end{figure}

In figure~\ref{fig8} we show the neighborhood of the transition region before and after the time
of maximum shear.  Four different collisions, separated by one million timesteps from their
predecessors and successors, are detailed there.  Between the open circles, the lack of pairing
is noticeable, and changes from one collision to the next.  Outside the transition region and
throughout, the reversed collision pairing is closely satisfied.  Let us next analyze the
contributions of the individual particles to the collision process within the transition region.

\begin{figure}[!t]
\centering
\includegraphics[height=0.45\textwidth,angle=-90.]{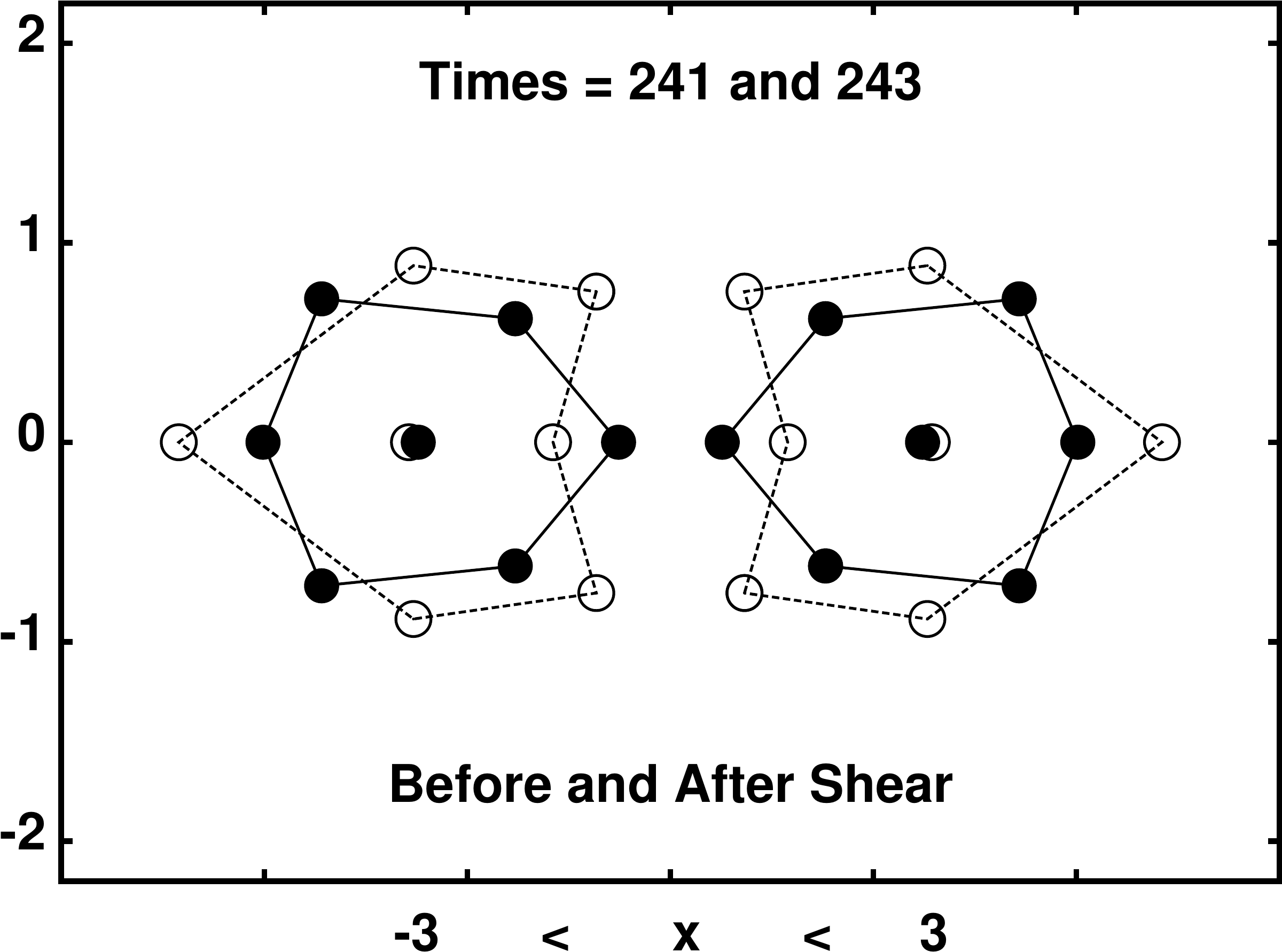} 
\caption{
Snapshots at times of 241 and 243, before (filled circles) and after (open circles) the
cooperative shearing motion of the central row of particles.
}\label{fig9}
\end{figure}

Figure~\ref{fig9} shows the individual-particle details of the change from the original hexagonal
crystallite shape during the collision.  In this low-speed example, two of the particles
in each hexagon shear over and under their neighbors in adjacent rows.

For each of the particles
we can quantify its importance to the dynamical instability by computing the set of all fourteen
particle amplitudes,
\[
\big\{ \delta x^2_i + \delta y^2_i + \delta {\dot x}^2_i + \delta \dot y^2_i  \big\}
 \longrightarrow \sum^N \delta^2_i \equiv 1 .
\]

\begin{figure}[!b]
\centering
\includegraphics[height=0.45\textwidth,angle=-90.]{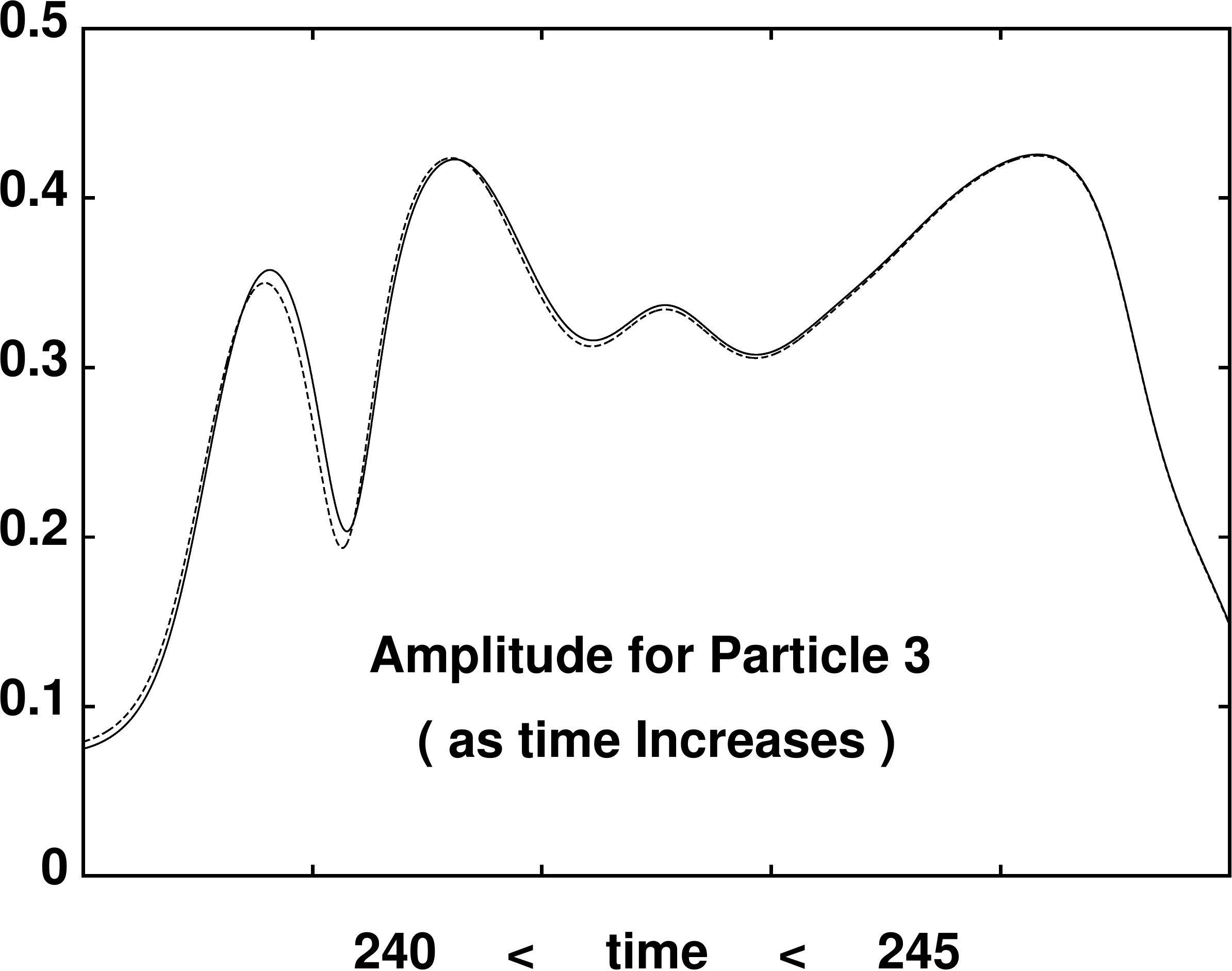}
\caption{
The contributions of the two ``important'' particles (the two filled circles closest to $x=0$ in
figure~\ref{fig9}) are identical and together account for about 90\% of the amplitude of $\delta_1^\textrm{f}$.  The
variation is shown in the time window of maximum shear,  $240 < t <  245$.  Data for the second
and third repetitions of the window are nearly identical and are shown as solid and dashed lines,
respectively. The initial velocities of the two 7-particle hexagons are $\dot x = \pm 0.1$.
}\label{fig10}
\end{figure}

Figure~\ref{fig10} shows the time-dependent amplitudes for Particles 3 and 10 during the maximum
shear period with time increasing: $240 < t < 245$.  The second and third repetitions of this
time period are both plotted here but nearly coincide.  Among the fourteen particles, Particle 3
(and its inverted image Particle 10) stand out with amplitudes near the maximum possible
(1/2) throughout the shearing motion.  Evidently, the lack of pairing during the collision
(as shown in figure~\ref{fig8}) is linked to the change in stability of the motions of these two particles as they pass by their neighbors, indicating the importance of shear to the stability of fluid deformation.

\begin{figure}[ht]
\centering
\includegraphics[width=0.66\textwidth]{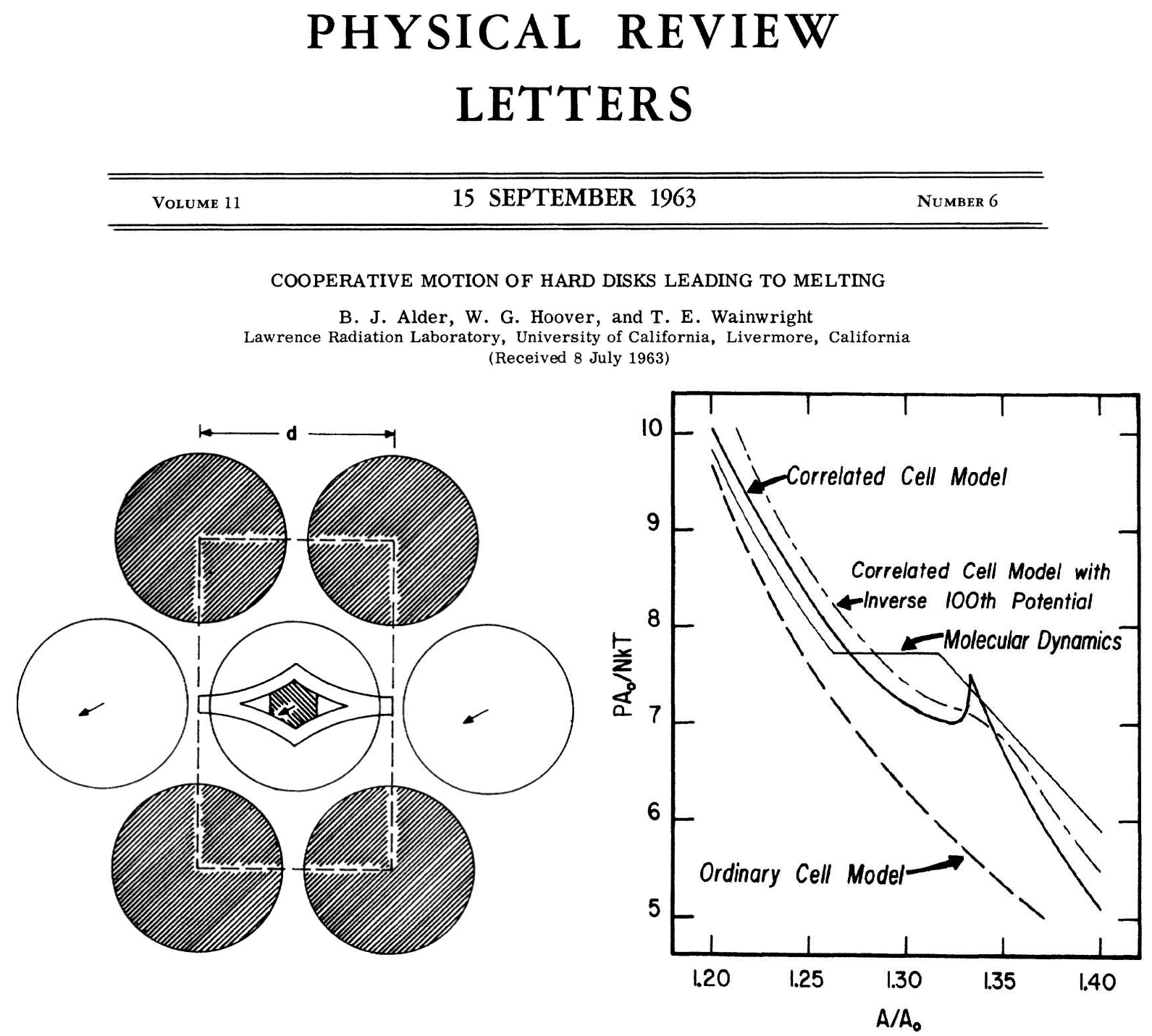}
\caption{
Figures from fifty years ago showing that the mechanism for hard-disk melting is the same as that
found for the irreversible aspect of an inelastic crystal collision.  The 2-particle van der
Waals loop at the right of the figure comes from the cooperative shearing motion shown at the
left for a periodic two-disk system. Reprinted from reference \cite{b18}.
}\label{fig11}
\end{figure}

{\it Fifty} years ago we helped point out that the hard-disk solid {\it melts} when it
becomes possible for shear of exactly this same kind to occur \cite{b18}.  See figure~\ref{fig11}.
This observation came from watching supercomputer movies of hard-disk molecular dynamics
at the Radiation Laboratory in Livermore.  The number of disks was 870 and the boundary
conditions were periodic.  Now this same mechanism can be seen anywhere in the world on a laptop computer.  It is interesting that despite all the change in computers and computation, the basic principles of mechanics, and the conclusions emerging from them, are unchanged.

\section{Conclusions}
\label{seq:7}

Many facets of classical mechanics still remain to be illuminated. Some of them directly
concern liquids and dynamic instability.  Today, computational advances make it possible to
evaluate Lyapunov spectra for systems with a few thousand exponents.  Just the first of
them, $\lambda_1(t)$, when compared to the last, $\lambda_n(t)$,  shows that the melting
mechanism identified from computer movies fifty years ago is still active today in the
irreversibly unstable collisional processes responsible for the Second Law of Thermodynamics.
The irreversibility of plastic flow in solids corresponds to the hysteresis of dislocation
motions.  In that flow process: stored energy $\longrightarrow$ heat.  The shearing motion
of dislocations is a close relative of the instability found in the present work.  Ergodicity
and the packing of hard disks has highlighted the importance of cooperative shearing to transitional
behavior \cite{b18,b19,b20}.  The ``free volumes'' accessible to liquid particles were of interest
in Doug Henderson and John Barker's work in 1976 \cite{b1}, in our work \cite{b19,b20}, and in
others' \cite{b21}.  The motivation for using periodic rather than rigid boundaries in molecular
dynamics simulations becomes very clear on examining the ``zoo'' of hard-disk structures which
can be ``locked in'' in the rigid case \cite{b21} where shear flow is prevented.

Nonequilibrium steady states generated with time-reversible thermostats \cite{b3} are relatively
simple to ``understand''.  When the long-time-averaged change in phase-space density is nonzero,
the only possibility consistent with stability (associated with a bounded phase-space distribution)
is a strange attractor, with the comoving change of density positive,  $\langle(\rd f/\rd t \rangle) > 0 $, while the fractal density itself is unchanged, $(\partial f/\partial t) \equiv 0$.  Familiarity
with this seemingly paradoxical state of affairs has led to its acceptance.  A similar
understanding of Hamiltonian irreversibility is not yet so clear.

An illuminating fringe benefit of our efforts was discovering that bit-reversible dynamics
may not retain the symmetry of its initial conditions.  Whether or not symmetry is maintained
is dependent upon the manner in which the particle forces are computed.  Retention cannot be
guaranteed unless the forces are summed up in an explicitly symmetric manner.
Long-time reversibility studies need to make this choice explicit.

We have emphasized that the local growth and coalescence rates in phase space are identical
going forward or backward in time \cite{b7}.  It is only through the influence of the ``past'' that
there can be a lack of symmetry in the local Lyapunov spectra.  This hidden source of
irreversibility is well worth mining as it is the most obvious property distinguishing one
direction of time travel from the other.
The lack of symmetry between the past and the future leaves its signature in the vectors
and the exponents identifying important particles.  The localization of these particles,
which jumps from spot to spot in larger systems \cite{b19}, is an explicitly time-irreversible
property, a tantalizing hint toward the understanding of Hamiltonian irreversibility in terms
of dynamical instabilities.

Though the pairing of exponents seems to be a widely-accepted consequence of Hamiltonian
mechanics, it is clear that the present results do violate pairing in the forward time direction.
In the reversed direction, fluctuations in pairing are at least two orders of magnitude smaller.
Why isn't pairing violated in the backward direction?  Evidently, there is still more work to be done.

\section*{Acknowledgment}

We thank the anonymous referee for several useful suggestions.  We address most of them here,
in order to clarify some of the underlying concepts and details of our work.

The referee  wished for evidence as to the Debye-like nature of manybody Lyapunov spectra.  This
became clear for manybody simulations during the 1980s.  In addition to figure~1 of reference~\cite{b6},
two nonequilibrium spectra appear as figure~1 in Wm.G.~Hoover, ``The Statistical Thermodynamics of Steady States'', Physics Letters A, 1999, \textbf{255}, 37--41.

That paper also illustrates the
fact that low-temperature ``stochastic'' thermostats, represented by a drag force $-(p/\tau)$,
can lead to fractal distributions.  This fractal nature, as judged by the Kaplan-Yorke dimension,
occurs because the phase-space offset vectors rotate very rapidly.  Vectors linking a reference
trajectory to an orthogonal array of rapidly-rotating satellite trajectories mix the
dimensionality loss associated with drag among {\it all} phase-space directions.

It needs to be
emphasized that the offset vectors are in phase space rather than tangent space.  The insensitivity
of our results to the length of these vectors and to the timestep was carefully checked.

The {\it inversion symmetry} of the balls and crystallites in figures~\ref{fig3}--\ref{fig5} implies that the top
of the leftmost projectile corresponds to the bottom of its rightmost image.  This inversion
symmetry is particularly noticeable in the ``important particles'' shown here in figure~\ref{fig3}.

As the referee kindly points out, it has been stated that Hamiltonian offset vectors are paired
so that the sum of the first and last local Lyapunov exponents should vanish exactly.  Here, we
simply state our finding that this pairing is sometimes violated, not only initially, but also
at long times.

The reason why we have used a collective density-dependent attractive potential for these
collision problems is that this reduces the yield strength so that the flow can more easily be
observed and analyzed.


\ukrainianpart

\title{Що таке рідина? Нестійкість Ляпунова розкриває необоротність порушення симетрії, яка прихована у багаточастинкових рівняннях руху Гамільтона}
 \author{В.Г. Гувер, К.Г. Гувер}
 \address{Дослідницький інститут м. Рубі Валей, Рубі Валей, Невада 89833, США
 }

\makeukrtitle

\begin{abstract}
 Типові рідини Гамільтона проявляють експоненціальну ``нестійкість Ляпунова'', яку також називають
 ``залежністю, чутливою до початкових умов''. Хоча рівняння Гамільтона є цілком оборотні по часу,
 прямі і зворотні необоротності Ляпунова можуть якісно різнитися. При числових обрахунках очікуване
 спарювання вперед/назад експонент Ляпунова також інколи порушується. Наприклад, розглянемо непружні
 зіткнення багатьох тіл у двовимірному просторі. Два дзеркально відображені зіштовхувані кристаліти
 можуть підстрибнути або ні, в результаті чого виникає єдина крапелька рідини, або декілька менших
 краплин в залежності від початкової кінетичної енегії та міжчастинкових сил. Відмінність між прямими
 і зворотніми еволюційними нестійкостями може корелювати з дисипацією та з другим законом термодинаміки.
 Отже, ці асиметричні стійкості рівнянь Гамільтона можуть забезпечити ``Стрілу Часу''. У даній статті ці
 факти проілюстровано для двох невеличких кристалітів, які зіштовхуються з тим, щоб утворити теплу рідину.
 Тут використано спеціально симетризовану форму біт-оборотнього інтегратора ``Ліп-фрог'' Левека і Верле.
 Проаналізовано траєкторії понад мільйони зіткнень при декількох рівновіддалених часових оберненнях.
\keywords нестійкість Ляпунова, експонентне спарювання, хаотична динаміка, необоротність
 %
 \end{abstract}

\end{document}